\def\ra{\rightarrow}
\def\prd#1#2#3{{\it Phys. Rev.} {\bf D#1} #2 (19#3)}
\def\pl#1#2#3{{\it Phys. Lett.} {\bf #1B} #2 (19#3)}
\def\np#1#2#3{{\it Nucl. Phys.} {\bf B#1} #2 (19#3)}
\def\prl#1#2#3{{\it Phys. Rev. Lett.} {\bf #1} #2 (19#3)}
\documentstyle[12pt,epsf]{article}
\def\beq{\begin{equation}}
\def\eeq{\end{equation}}

\def\beqn{\begin{eqnarray}}
\def\eeqn{\end{eqnarray}}
\relax
\begin{document}
\begin{titlepage}
\def\ba{\begin{array}}
\def\ea{\end{array}}
\def\thefootnote{\fnsymbol{footnote}}
\vfill
\hskip 4in BNL-60949

\hskip 4in October, 1994
\vspace{1 in}
\begin{center}
{\large \bf BOUNDS ON ANOMALOUS GAUGE BOSON COUPLINGS
FROM PARTIAL $Z$ WIDTHS AT LEP}\\

\vspace{1 in}
{\bf S.~Dawson$^{(a)}$}\footnote{This manuscript has been authored
under contract number DE-AC02-76CH00016 with the U.S. Department
of Energy.  Accordingly, the
U.S. Government retains a non-exclusive, royalty-free license to
publish or reproduce the published form of this contribution, or
allow others to do so, for U.S. Government purposes.}
{\bf  and G.~Valencia$^{(b)}$}\\
{\it  $^{(a)}$ Physics Department,
               Brookhaven National Laboratory,  Upton, NY 11973}\\
{\it  $^{(b)}$ Department of Physics,
               Iowa State University,
               Ames IA 50011}\\
\vspace{1 in}
\end{center}
\begin{abstract}
We place bounds on anomalous gauge boson couplings from LEP data
with particular emphasis on those couplings which do not contribute
to $Z$ decays at tree level. We use an effective field theory formalism
to compute the one-loop corrections to the $Z\rightarrow {\overline f} f$
decay widths resulting from non-standard model three and four gauge boson
vertices. We find that the precise measurements at LEP constrain the three
gauge boson couplings at a level comparable to that obtainable at LEPII
and LHC.
\end{abstract}

\end{titlepage}

\clearpage

\section{Introduction}

High precision measurements at the $Z$ pole at LEP
combined with polarized forward backward asymmetries at SLC
and other measurements of electroweak observables at lower energies
have been used to place stringent limits on new physics beyond the standard
model \cite{altanew,langanew}.

Under the assumption that the dominant effects of the new physics
would show up as corrections to the gauge boson self-energies, the
LEP measurements have been used to parameterize the possible new
physics in terms of three observables $S$, $T$, $U$
\cite{pestak}; or equivalently
$\epsilon_1$, $\epsilon_2$, $\epsilon_3$ \cite{alta}. The difference
between the two
parameterizations is the reference point which corresponds to the
standard model predictions. A fourth observable
corresponding to the partial width $Z \ra b \overline{b}$ has
been analyzed in terms of the parameter $\delta_{bb}$ \cite{langanew} or
$\epsilon_b$ \cite{alta}.

In view of the extraordinary agreement between the standard model
predictions and the observations, it seems reasonable to assume
that the $SU(2)_L \times U(1)_Y$ gauge theory of electroweak interactions
is essentially correct, and that the only sector of the theory
lacking experimental support is the symmetry breaking sector. There
are many extensions of the minimal standard model that incorporate
different symmetry breaking possibilities. One large class of models
is that in which the interactions responsible for the symmetry breaking
are strongly coupled. For this class of models one expects that there
will be no new particles with masses below 1~TeV or so, and that their
effects would show up in experiments as deviations from the minimal
standard model couplings.

In this paper we use the latest measurements of partial decay widths
of the $Z$ boson to place bounds on anomalous gauge boson couplings.
Our paper is organized as follows. In Section~2 we discuss our formalism
and the assumptions that go into the relations between the partial
widths of the $Z$ boson and the anomalous couplings. In Section~3
we present our results. Finally, in Section~4 we discuss the
difference between our calculation and others that can be found in
the literature, and assess the significance of our results by comparing
them to other existing limits. Detailed analytical
formulae for our results are relegated to an appendix.

\section{Formalism}

We assume that the electroweak interactions are given
by an $SU(2)_L \times U(1)_Y$ gauge
theory with spontaneous symmetry breaking to $U(1)_{EM}$, and
that we do not have any information on the symmetry breaking sector
except that it is strongly interacting and that any new particles have
masses higher than several hundred GeV. It is well known that this
scenario can be described with an effective Lagrangian with operators
organized according to the number of derivatives or gauge fields
they have. If we call $\Lambda$ the scale at which the symmetry breaking
physics comes in, this organization of operators corresponds to
an expansion of amplitudes in powers of $(E^2~{\rm or}~v^2)/\Lambda^2$.
For energies $E \leq v$ this is just an expansion in powers of the
coupling constant $g$ or $g^\prime$, and for energies $E \geq v$ it
becomes an energy expansion.
The lowest order effective Lagrangian for the symmetry breaking
sector of the theory is \cite{longo}:
\beq
{\cal L}^{(2)}={v^2 \over 4}{\rm Tr}\biggl[D^\mu \Sigma^\dagger D_\mu
\Sigma \biggr].
\label{lagt}
\eeq
In our notation $W_{\mu}$ and $B_{\mu}$ are
the $SU(2)_L$ and $U(1)_Y$  gauge fields with
$W_\mu \equiv W^i_\mu \tau_i$.\footnote{$Tr(\tau_i \tau_j)=2\delta_{ij}$.}
The matrix $\Sigma \equiv \exp(i\vec{\omega}\cdot \vec{\tau} /v)$, contains the
would-be Goldstone bosons $\omega_i$ that give the $W$ and $Z$ their
masses via the Higgs mechanism and the $SU(2)_L \times U(1)_Y$
covariant derivative is given by:
\beq
D_\mu \Sigma = \partial_\mu \Sigma +{i \over 2}g W_\mu^i \tau^i\Sigma
-{i \over 2}g^\prime B_\mu \Sigma \tau_3.
\label{covd}
\eeq
Eq.~\ref{lagt} is thus the $SU(2)_L \times U(1)_Y$ gauge
invariant mass term for the $W$ and $Z$. The physical masses
are obtained with $v \approx 246$~GeV. This non-renormalizable Lagrangian
is interpreted as an effective field theory, valid below
some scale $\Lambda \leq 3$~TeV. The lowest order interactions between
the gauge bosons and fermions, as well as the kinetic energy
terms for all fields,  are the same as those in the minimal
standard model.

For LEP observables, the operators that can appear at tree-level
are those that modify the gauge boson self-energies. To order
${\cal O}(1/\Lambda^2)$ there are only three \cite{longo,appel}:
\beq
{\cal L}^{(2GB)}=\beta_1{v^2\over 4} \biggl({\rm Tr} \biggl[
\tau_3 \Sigma^\dagger D_\mu \Sigma\biggr]\biggr)^2
+ \alpha_8g^2 \biggl({\rm Tr}\biggl[\Sigma \tau_3 \Sigma^\dagger
W_{\mu \nu}\biggr]\biggr)^2
 + g g^{\prime}{v^2 \over \Lambda^2} L_{10}\, {\rm Tr}\biggl[ \Sigma
B^{\mu \nu}
\Sigma^\dagger W_{\mu \nu}\biggr],
\label{oblique}
\eeq
which contribute respectively to $T$, $U$ and $S$. Notice that for the
two operators that break the custodial $SU(2)_C$ symmetry we have
used the notation of Ref.~\cite{longo,appel}.

In this paper we will consider operators that affect the $Z$ partial
widths at the one-loop level. We will restrict our study to only those
operators that appear at order ${\cal O}(1/\Lambda^2)$ in the gauge-boson
sector and that respect the custodial symmetry in the limit
$g^\prime \ra 0$. They are:
\beqn
{\cal L}^{(4)}\ &=&\ {v^2 \over \Lambda^2}  \biggl\{ L_1 \,
\biggl( {\rm Tr}\biggl[
D^\mu\Sigma^\dagger D_\mu \Sigma \biggr]\biggl)^2
\ +\  L_2 \,\biggl({\rm Tr}\biggl[
 D_\mu\Sigma^\dagger D_\nu \Sigma\biggr]\biggl)^2
 \nonumber \\
&& - i g L_{9L} \,{\rm Tr}\biggl[W^{\mu\nu} D_\mu
\Sigma D_\nu \Sigma^\dagger\biggr]
\ -\ i g^{\prime} L_{9R} \,
{\rm Tr}\biggl[ B^{\mu \nu}
D_\mu \Sigma^\dagger D_\nu\Sigma\biggr]\biggr\},
\label{lfour}
\eeqn
where the field strength tensors are given by:
\beqn
W_{\mu\nu}&=&{1 \over 2}\biggl(\partial_\mu W_\nu -
\partial_\nu W_\mu + {i \over 2}g[W_\mu, W_\nu]\biggr)
\nonumber \\
B_{\mu\nu}&=&{1\over 2}\biggl(\partial_\mu B_\nu-\partial_\nu B_\mu\biggr)
\tau_3.
\label{fsten}
\eeqn
Although this is not a complete list of all the operators that can arise
at this order, we will be able to present a consistent picture in
the sense that our calculation will not require additional counterterms
to render the one-loop results finite. Our choice of this subset of operators
is motivated by the theoretical prejudice that violation of custodial
symmetry must be ``small'' in some sense in the full theory \cite{siki}.
We want to restrict our attention to a small subset of all the operators that
appear at this order because there are only a few observables that have been
measured.

The operators in Eq.~\ref{oblique} and Eq.~\ref{lfour} would
arise when considering the effects
of those in Eq.~\ref{lagt} at the one-loop level, or
from the new physics responsible for symmetry
breaking at a scale $\Lambda$ at order $1/\Lambda^2$. We, therefore,
explicitly introduce the factor $v^2/\Lambda^2$ in our definition
of ${\cal L}^{(4)}$ so that the coefficients
$L_i$ are naturally of ${\cal O}(1)$.\footnote{
We do not introduce this factor in
the $SU(2)_C$ violating couplings $\beta_1$ and $\alpha_8$ since we
do not concern ourselves with them in this paper. They are simply
used as counterterms for our one-loop calculation.}

The anomalous couplings that we consider would have tree-level effects
on some observables that can be studied in future colliders. They have been
studied at length in the literature \cite{boud}.
In the present paper we will compute their
contribution to the $Z$ partial widths that are measured at LEP. These
operators contribute to the $Z$ partial widths
at the one-loop level. Since we
are dealing with a non-renormalizable effective Lagrangian, we will
interpret our one-loop results in the usual way of an effective field theory.

We will first perform a complete calculation to order ${\cal O}
(1/\Lambda^2)$. That is, we will include the one-loop contributions from
the operator in Eq.~\ref{lagt} (and gauge boson kinetic energies).
The divergences generated in this calculation are absorbed by
renormalization of the couplings in Eq.~\ref{oblique}. This calculation
will illustrate our method, and as an example, we use it to place
bounds on $L_{10}$.

We will then place bounds on the couplings of Eq.~\ref{lfour} by considering
their one-loop effects. The divergences generated in this one-loop calculation
would be removed in general by renormalization of the couplings in the
${\cal O}(1/\Lambda^4)$ Lagrangian of those operators that modify the
gauge boson self-energies at tree-level; and perhaps by additional
renormalization of the couplings in Eq.~\ref{oblique}. This would occur
in a manner analogous to our ${\cal O}(1/\Lambda^2)$ calculation.
Interestingly,
we find that we can obtain a completely finite result for the $Z \ra
\overline{f} f$ partial widths using only the operators in Eq.~\ref{oblique}
as counterterms.

However, our interest is to place bounds on the couplings
of Eq.~\ref{lfour} so we proceed as follows. We first regularize the
integrals in $n$ space-time dimensions and remove all the poles in
$n-4$ as well as the finite analytic terms by a suitable definition of
the renormalized couplings. We then base our
analysis on the leading non-analytic
terms proportional to $L_i \log \mu$. These terms determine the
running of the $1/\Lambda^4$ couplings and cannot be generated by
tree-level terms at that order. It has been argued in the literature
\cite{georgi}, that with a carefully chosen renormalization scale $\mu$
(in such a way that the logarithm is of order one), these terms give us
the correct order of magnitude for the size of the $1/\Lambda^4$
coefficients. We thus choose some value for the renormalization
scale between the $Z$ mass and $\Lambda$ and require that this
logarithmic contribution to the renormalized couplings falls in the
experimentally allowed range. Clearly, the LEP observables do not
measure the couplings in Eq.~\ref{lfour}, and it is only from
naturalness arguments like the one above, that we can place bounds
on the anomalous gauge-boson couplings.
{}From this perspective, it is clear that these bounds are not a substitute
for direct measurements in future high energy machines. They should,
however, give us an indication for the level at which we can expect
something new to show up in those future machines.

We will perform our calculations in unitary gauge, so we set $\Sigma=1$
in Eqs.~\ref{lagt}, \ref{oblique} and \ref{lfour}. This results in interactions
involving three, and four gauge boson couplings, some of which we
present in Appendix~A. Those
coming from Eq.~\ref{lagt} are equivalent to those in the minimal
standard model with an infinitely heavy Higgs boson, and those coming
from Eq.~\ref{lfour} correspond to the ``anomalous'' couplings.

For the lowest order operators we use the conventional
input parameters: $G_F$ as measured in muon decay;
the physical $Z$ mass: $M_Z$; and $\alpha(M_Z)=1/128.8$. Other lowest
order parameters are derived quantities and we adopt one of the
usual definitions for the mixing angle:
\beq
s_Z^2 c_Z^2 \equiv{\pi \alpha(M_Z)\over \sqrt{2} G_F M_Z^2}.
\label{szdef}
\eeq

We neglect the mass and momentum of the external fermions
compared to the $Z$ mass. In particular, we do not include the $b$-quark
mass since it would simply introduce corrections of order
$5\%$ and our results are only order of magnitude estimates.
The only fermion mass that is kept in our calculation is
the mass of the top-quark when it appears as an intermediate state.

With this formalism we proceed to compute the $Z\ra f \overline{f}$
partial width from the following ingredients.

\begin{itemize}

\item The $Z\ra f {\overline f}$ vertex, which we write as:
\beq
i {\Gamma}_\mu =  -i{e\over 4 s_Z c_Z}
\gamma_\mu\biggl[
(r_f +\delta r_f)(1+\gamma_5) +(l_f+\delta l_f)
 (1-\gamma_5)\biggr ]
\label{vertex}
\eeq
where $r_f=-2Q_f  s_Z^2$ and $l_f=r_f+T_{3f}$.
The terms $\delta l_f$ and $\delta r_f$ occur at one-loop both at order
$1/\Lambda^2$ and at order $1/\Lambda^4$ and are given in Appendix~B.

\item The renormalization of the lowest order input parameters.
At order $1/\Lambda^2$ it is induced by tree-level anomalous
couplings and one-loop diagrams with lowest order vertices. At
order $1/\Lambda^4$ it is induced by one-loop diagrams with an
anomalous coupling in one vertex. We present analytic formulae
for the self-energies, vertex corrections and boxes in Appendix~B.
The changes induced in the lowest order input parameters are:
\beqn
{\Delta \alpha\over \alpha} &=& {A_{\gamma \gamma}(q^2)\over q^2}
\mid _{q^2=0}
\nonumber \\
{\Delta M_Z^2\over M_Z^2}&=&{A_{ZZ}(M_Z^2)\over M_Z^2}
\nonumber \\
 {\Delta G_F\over G_F} & = & {2 \Gamma_{W e \nu}\over {\cal A}_0}
-{A_{WW}(0)\over M_W^2}+(Z_f-1)+B_{\rm box}
\label{shifts}
\eeqn
The self-energies $A_{VV}$ receive tree-level contributions from the
operators with $L_{10}$, $\beta_1$ and $\alpha_8$. They also
receive one-loop contributions from the lowest
order Lagrangian Eq.~\ref{lagt} and from the operators with
$L_1$, $L_2$, $L_{9L}$ and $L_{9R}$. The effective $W e \nu$ vertex,
$\Gamma_{W e \nu}$, receives one-loop contributions from all
the operators in Eq.~\ref{lfour}. The fermion wave function
renormalization factors
$Z_f$ and the box contribution to $\mu \rightarrow e \nu
{\overline \nu}$, $B_{\rm box}$, are due only to one-loop
effects from the lowest order effective Lagrangian and are thus
independent of the anomalous couplings. Notice that $B_{\rm box}$ enters the
renormalization of $G_F$ because we work in unitary gauge where box
diagrams also contain divergences.

\item Tree-level and one-loop contributions to $\gamma Z$ mixing.
Instead of diagonalizing the neutral gauge boson
sector, we include this mixing as an additional
contribution to $\delta l_f$ and $\delta r_f$ in Eq.~\ref{vertex}:
\beq
\delta l_f^{\prime} = \delta r_f^{\prime}
 =  -{ c_Z \over s_Z} r_f { A_{\gamma Z}(M_Z^2)  \over M_Z^2}
\label{mix}
\eeq

\item Wave function renormalization. For the external fermions we include it
as additional contributions to $\delta l_f$ and $\delta r_f$ as shown in
Appendix~B. For the $Z$ we include it explicitly.

\end{itemize}

With all these ingredients we can collect the results from Appendix~B into
our final expression for the physical partial width. We find:
\beqn
\Gamma(Z\rightarrow f {\overline f}) &
=& \Gamma_0 Z_Z \biggl[ 1
-{\Delta G_f \over G_f} - {\Delta M_Z^2 \over M^2_Z}
+{2(l_f \delta l_f + r_f \delta r_f)\over l_f^2 + r_f^2}
\nonumber \\ && -
{2 r_f(l_f + r_f)\over l_f^2 + r_f^2}{c^2_Z \over s^2_Z -c^2_Z}
\biggl({\Delta G_f \over G_f} + {\Delta M_Z^2 \over M^2_Z}
-{\Delta \alpha \over \alpha} \biggr)\biggr],
\label{full}
\eeqn
where $\Gamma_0$ is the lowest order tree level result,
\beq
\Gamma_0 (Z\rightarrow f {\overline f})=
N_{cf} (l_f^2 + r_f^2) {G_F M_Z^3 \over 12 \pi \sqrt{2}},
\label{widthl}
\eeq
and $N_{cf}$ is 3 for quarks and 1 for leptons.
We write the contributions of the different anomalous couplings
to the $Z$ partial widths in the form:
\beq
\Gamma(Z\ra f \overline{f}) \equiv \Gamma_{SM}(Z\ra f\overline{f})
\biggl( 1 + {\delta \Gamma_f^{L_i} \over \Gamma_0(Z\ra f\overline{f})}\biggr).
\label{defw}
\eeq
We use this form because we want to place bounds on the anomalous
couplings by comparing the measured widths with the one-loop standard
model prediction $\Gamma_{SM}$. Using Eq.~\ref{defw} we introduce
additional terms proportional to products of standard model one-loop
corrections and corrections due to anomalous couplings. These are
small effects that do not affect our results.

We will not attempt to obtain a global fit to the parameters in our
formalism from all possible observables. Instead we use the
partial $Z$ widths. We believe this approach to be adequate
given the fact that the results  rely on naturalness assumptions.
Specifically we consider the observables:
\beqn
\Gamma_e &=&  83.98\pm 0.18 ~{\rm MeV}~{\rm~Ref.~\cite{altanew}}
\nonumber \\
\Gamma_\nu &=& 499.8 \pm 3.5~{\rm MeV}~{\rm~Ref.~\cite{glas}}
\nonumber \\
\Gamma_Z &=&2497.4 \pm 3.8~{\rm MeV}~{\rm~Ref.~\cite{glas}}
\nonumber \\
R_h &=& 20.795\pm 0.040~{\rm~Ref.~\cite{glas}}
\nonumber \\
R_b &=& 0.2202\pm 0.0020~{\rm~Ref.~\cite{glas}}
\label{data}
\eeqn
The bounds on new physics are obtained by subtracting the standard
model predictions at one-loop from the measured partial widths
as in Eq.~\ref{defw}. We
use the numbers of Langacker \cite{langanew} which use the global
best fit values for $M_t$ and $\alpha_s$ with $M_H$ in the range
$60-1000$~GeV. The first error is from the uncertainty in $M_Z$ and
$\Delta r$, the second is from $M_t$ and $M_H$, and the one in
brackets is from the uncertainty in $\alpha_s$.
\beqn
\Gamma_e &=& 83.87~\pm 0.02 \pm 0.10~{\rm MeV}~{\rm~Ref.~\cite{langap}}
\nonumber \\
\Gamma_\nu &=& 501.9 \pm 0.1 \pm 0.9~{\rm MeV}~{\rm~Ref.~\cite{bernabeu}}
\nonumber \\
\Gamma_Z &=& 2496 \pm 1 \pm 3 \pm [3]~{\rm MeV}~{\rm~Ref.~\cite{langanew}}
\nonumber \\
R_h &=& 20.782 \pm 0.006 \pm 0.004 \pm[0.03]~{\rm~Ref.~\cite{langanew}}
\nonumber \\
\delta_{bb}^{new} &=& 0.022 \pm 0.011~{\rm~Ref.~\cite{langap}}
\label{theory}
\eeqn
where $\delta_{bb}^{new}\equiv [\Gamma(Z \ra b\overline{b})
-\Gamma(Z \ra b\overline{b})^{(SM)}]/\Gamma(Z \ra b\overline{b})^{(SM)}$.
We add all errors in quadrature.

\section{Results}

In this section we compute the corrections to the $Z \ra f \overline{f}$
partial widths from the couplings of Eq.~\ref{lfour}, and compare them
to recent values measured at LEP. We treat each coupling constant
independently, and compute only its {\it lowest} order contribution to the
decay widths. We first present the complete ${\cal O}(1/\Lambda^2)$
results. They illustrate our method and serve as a check of our calculation.
We then look at the effect of the couplings $L_{1,2}$ which affect only
the gauge-boson self-energies. We then study the more complicated case of the
couplings $L_{9L,9R}$. Finally we isolate the non-universal effects
proportional to $M_t^2$. As explained in the previous section, we do not
include in our analysis the operators that appear at ${\cal O}(1/\Lambda^2)$
that break the custodial symmetry. As long as one is interested in
bounding the anomalous couplings one at a time, it is straightforward
to include these operators. For example, we discussed the parity violating
one in Ref.~\cite{dvplep}.

\subsection{Bounds on $L_{10}$ at order $1/\Lambda^2$.}

The operators in Eq.~\ref{oblique} are the only ones that induce a
tree-level correction to the gauge boson self-energies to
order ${\cal O}(1/\Lambda^2)$. This can be
seen most easily by working in a physical basis in which the neutral gauge
boson self-energies are diagonalized to order ${\cal O}(1/\Lambda^2)$.
This is accomplished with renormalizations described in the literature
\cite{holdom,cdhv}, and results in modifications to
the $W {\overline f}^{\prime} f$ and the $Z f \overline{f}$ couplings.
This tree-level effect on the $Z \ra f \overline{f}$ partial
width is, of course, well known. It corresponds,
{\it at leading order}, to the new physics
contributions to $S,~T,~U$ or $\epsilon_{1,2,3}$
discussed in the literature \cite{alls}.

In this section we do not perform the diagonalization mentioned above, but
rather work in the original basis for the fields. This will serve two
purposes. It will allow us to present a complete ${\cal O}(1/\Lambda^2)$
calculation as an illustration of the method we use to bound the other
couplings. Also, because the gauge boson interactions that appear at
this order have the same tensor structure as those induced by
$L_{9L}$ and $L_{9R}$, we will be able to carry out the calculation
involving those two couplings simultaneously. In this way, even though
the terms with $L_{9L,9R}$ are order $1/\Lambda^4$,
the calculation to order $1/\Lambda^2$ will serve as a check of
our answer for $L_{9L,9R}$.

To recover the $1/\Lambda^2$ result we set $L_{9L,9R}=0$ (and also
$L_{1,2}=0$ but these terms are clearly different) in the results
of Appendix~B. As explained in Section~2, we have regularized our
one-loop integrals in $n$ dimensions and isolated the  ultraviolet
poles $1/\epsilon = 2/(4-n)$. We find that we obtain a finite answer
to order $1/\Lambda^2$ if we adopt the following renormalization
scheme:\footnote{That is, a finite answer for the physical observables
$Z \ra \overline{f} f$, not for quantities like the self-energies.}
\footnote{
In these expressions we have simply dropped any finite constants
arising from the loop calculation. These constants would only
be of interest in a complete effective field theory analysis
applied to a problem where all unknown constants can be measured
and additional predictions made. For example, that is the status of
the ${\cal O}(p^4)$ $\chi$PT theory for low energy strong interactions.}
\beqn
{v^2 \over \Lambda^2} L^r_{10}(\mu) &=& {v^2 \over \Lambda^2}
L_{10}-{1 \over 16 \pi^2} {1 \over 12}
\biggl( {1 \over \epsilon} + \log{\mu^2 \over M_Z^2} \biggr)
\nonumber \\
\beta_1^r(\mu) &=& \beta_1 -
{e^2 \over 16 \pi^2}{3 \over 2 c^2_Z}
\biggl( {1 \over \epsilon} + \log{\mu^2 \over M_Z^2} \biggr).
\label{renorm}
\eeqn
We thus replace the bare parameters $L_{10}$ and $\beta_1$ with the
scale dependent ones above. As a check of our answer, it is interesting
to note that we would also obtain a finite answer by adding to the
results of Appendix~B, the one-loop contributions to the self-energies
obtained in unitary gauge in the minimal standard model with one Higgs
boson in the loop. Equivalently, the expressions in Eq.~\ref{renorm}
correspond to the value of $L_{10}$ and $\beta_1$ at one-loop in the
minimal standard model within our renormalization scheme.\footnote{Our
result agrees with that of Refs.~\cite{longo,herrero}.}

Our result for $L_{10}$ at order $1/\Lambda^2$ is then:
\beq
{\delta \Gamma_f^{L_{10}} \over \Gamma_0(Z\ra f\overline{f})}=
{e^2 \over c_Z^2 s_Z^2}L^r_{10}(\mu)
{v^2 \over \Lambda^2}{2r_f(l_f +r_f) \over
l_f^2+r_f^2}{c_Z^2 \over s_Z^2-c_Z^2}
\label{shten}
\eeq
Once again we point out that, at this order, the contribution of
$L_{10}$ to the
LEP observables occurs only through modifications to the self-energies
that are proportional to $q^2$. At this order it is therefore
possible to identify the effect of $L_{10}$ with the oblique
parameter $S$ or $\epsilon_3$. If we were to compute the effects
of $L_{10}$ at one-loop (as we do for the $L_{9L,9R}$), comparison with
$S$ would not be appropriate. Bounding $L_{10}$ from existing analyses
of $S$ or $\epsilon_3$ is complicated by the fact that the
same one-loop definitions must be used. For example, $L_{10}(\mu)$ receives
contributions from the standard model Higgs boson that are usually
included in the minimal standard model calculation. We will simply
associate our definition of $L_{10}(\mu)$ with {\it new} contributions
to $S$, beyond those coming from the minimal stardard
model.\footnote{We are being sloppy by not matching our $L_{10}$ at
one loop with the precise definitions used to renormalize the
standard model at one-loop. This does not matter for our present
purpose.}

Numerically we find the following $90\%$ confidence level bounds on $L_{10}$
when we take the scale $\Lambda=2$~TeV:
\beqn
\Gamma_e &\ra & -1.7 \leq L^r_{10}(M_Z)_{new} \leq 3.3 \nonumber \\
R_h &\ra & -1.5 \leq L^r_{10}(M_Z)_{new} \leq 2.0 \nonumber \\
\Gamma_Z &\ra & -1.1 \leq L^r_{10}(M_Z)_{new} \leq 1.5
\label{tennum}
\eeqn
We can also bound the {\it leading order} effects of
$L_{10}$ from Altarelli's latest global fit
$\epsilon_3 = (3.4 \pm 1.8) \times 10^{-3}$ \cite{altanew}. To do this,
we subtract the standard model value obtained with $160\leq M_t \leq
190$~GeV and $65 \leq M_H \leq 1000$~GeV as read from Fig.~8 in
Ref.~\cite{altanew}. We obtain the $90\%$ confidence level interval:
\beq
-0.14 \leq L^r_{10}(M_Z)_{new} \leq 0.86
\label{sfalta}
\eeq
We can also compare directly with the result of Langacker
$S_{new}=-0.15\pm 0.25^{-0.08}_{+0.17}$\cite{langanew}, to obtain
$90\%$ confidence level limits:
\beq
-0.46 \leq L^r_{10}(M_Z)_{new} \leq 0.77
\label{sflanga}
\eeq
The results Eqs.\ref{sfalta}, \ref{sflanga} are better than our result
Eq.~\ref{tennum} because they correspond to global fits that include
all observables.

\subsection{Bounds on $L_{1,2}$ at order $1/\Lambda^4$.}

The couplings $L_{1,2}$ enter the one-loop
calculation of the $Z \ra f \overline{f}$
width through four gauge boson couplings as depicted schematically
in Figure~1.
\begin{figure}[htb]
\centerline{\hfil\epsffile{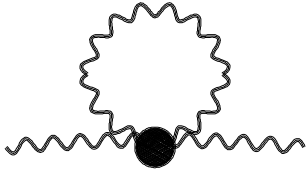}\hfil}
\caption[]{Gauge boson self-energy diagrams involving the couplings $L_{1,2}$.}
\end{figure}
Our prescription
calls for using only the leading non-analytic contribution
to the process $Z \ra f \overline{f}$. This contribution can be extracted
from the coefficient of the pole in $n-4$.
Care must be taken to isolate the poles of
ultraviolet origin (which are the only ones that interest us) from those
of infrared origin that appear in intermediate steps of the calculation
but that cancel as usual when one includes real emission processes as well.
We thus use the results of Appendix~B with the replacement:
\beq
{1 \over \epsilon} = {2 \over 4-n} \ra \log\biggl({\mu^2 \over M_Z^2} \biggr)
\label{prescrip}
\eeq
to compute the contributions to the partial widths using Eq.~\ref{full}.

Since in unitary gauge $L_{1,2}$ modify only the four-gauge
boson couplings at the one-loop level, they
enter the calculation of the $Z$ partial widths only through the
self-energy corrections and Eq.~\ref{shifts}.
These operators induce a non-zero value for $\Delta \rho \equiv
\Pi_{WW}(0) / M_W^2 - \Pi_{ZZ}(0)/ M_Z^2$. For the observables
we are discussing, this is the {\it only} effect of $L_{1,2}$.
We do not place bounds on them from global fits of the oblique
parameter $T$ or $\epsilon_1$, because we have not shown that
this is the only effect of $L_{1,2}$ for the other observables
that enter the global fits. It is curious to see that even
though the operators with $L_1$ and $L_2$ violate the custodial
$SU(2)_C$ symmetry {\it only} through the hypercharge coupling,
their one-loop effect on the partial $Z$ widths is equivalent to
a $g^4$ contribution to $\Delta\rho$, on the same footing as two-loop
electroweak contributions to $\Delta\rho$ in the minimal standard
model. The calculation to ${\cal O}(1/\Lambda^4)$
can be made finite with the following renormalization of $\beta_1$:
\beq
\beta_1^r(\mu) = \beta_1
+{3 \over 4}{\alpha^2 (1 + c_Z^2)\over s_Z^2 c_Z^4}
\biggl(L_1 + {5 \over 2} L_2\biggr){v^2 \over \Lambda^2}
\biggl({1 \over \epsilon}+\log{\mu^2 \over M_Z^2}\biggr).
\label{shone}
\eeq

Using our prescription to bound the anomalous couplings, Eq.~\ref{prescrip},
we obtain for the $Z$ partial widths:
\beq
{\delta\Gamma_f^{L_{1,2}} \over \Gamma_0(Z\ra f \overline{f})}=
-{3 \over 2}{\alpha^2 (1 + c_Z^2)\over s_Z^2 c_Z^4}
\biggl(L_1 + {5 \over 2} L_2\biggr){v^2 \over \Lambda^2}\log\biggl(
{\mu^2 \over M_Z^2}\biggr)
\biggl(1 +{2 r_f (l_f + r_f) \over l_f^2 + r_f^2}{c_Z^2 \over s_Z^2 - c_Z^2}
\biggr).
\label{resonetwo}
\eeq
Using $\Lambda=2$~TeV, and $\mu=1$~TeV we find $90\%$ confidence level bounds:
\beqn
\Gamma_e &\ra & -50 \leq L_1 + {5 \over 2}L_2  \leq 26 \nonumber \\
\Gamma_\nu &\ra & -28 \leq L_1 + {5 \over 2}L_2  \leq 59 \nonumber \\
R_h &\ra & -190 \leq L_1 + {5 \over 2}L_2 \leq 130 \nonumber \\
\Gamma_Z &\ra & -36 \leq L_1 + {5 \over 2}L_2 \leq 27
\label{rhoana}
\eeqn
Combined, they yield the result:
\beq
-28 \leq L_1 + {5 \over 2} L_2 \leq 26
\label{combineonetwo}
\eeq
shown in Figure~2.
\begin{figure}[htb]
\centerline{\hfil\epsffile{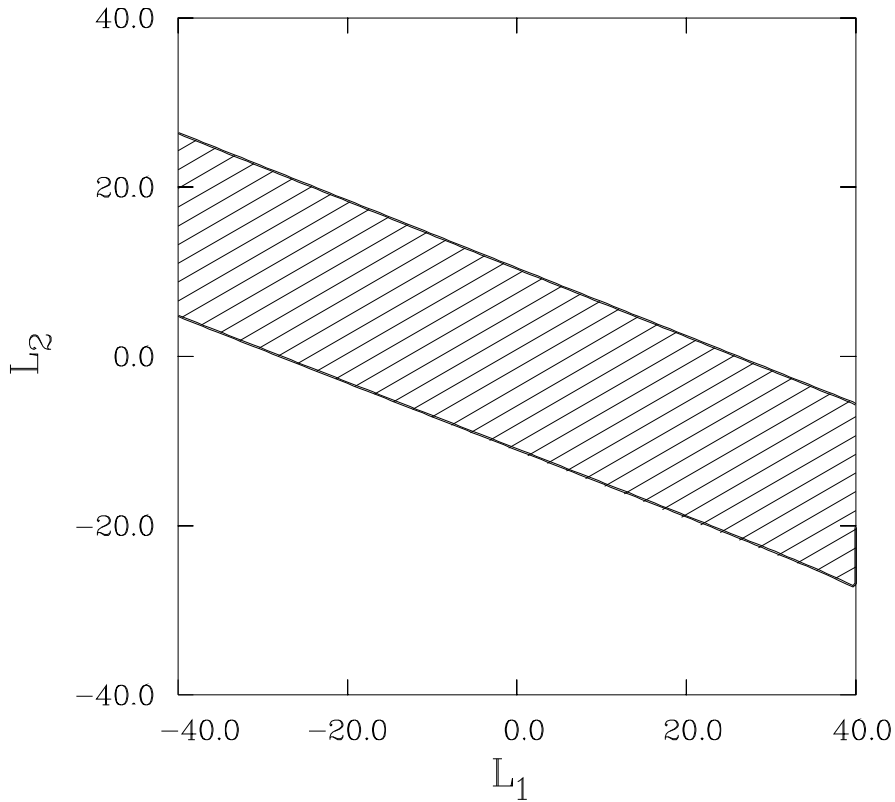}\hfil}
\caption[]{$90\%$ confidence level bounds on $L_{1,2}$  from the $Z\rightarrow
f {\overline f}$ partial widths, (Eq. 24).  The allowed region is shaded.}
\end{figure}
As mentioned before, the effect of $L_{1,2}$ in
other observables is very different from that of $\beta_1$.\footnote{
An example are the observables discussed by us in Ref.\cite{bdv}.} It is
only for the $Z$ partial widths that we can make the ${\cal O}
(1/\Lambda^4)$ calculation finite with Eq.~\ref{shone}.

\subsection{Bounds on $L_{9L,9R}$ at order $1/\Lambda^4$.}

The operators with $L_{9L}$ and $L_{9R}$ affect the
$Z$ partial widths through Eqs.~\ref{vertex}, \ref{shifts}, and
\ref{mix}. We find it convenient to carry out this
calculation simultaneously with the one-loop effects of the lowest
order effective Lagrangian, Eq.~\ref{lagt}, because
the form of the three and four gauge
boson vertices induced by these two couplings is the same as that
arising from Eq.~\ref{lagt}. This can be seen from Eqs.~\ref{conv},~\ref{unot}
in Appendix~A. Performing the calculation in this way, we obtain a result
that contains terms of order $1/\Lambda^2$ (those independent of $L_{9L,9R}$),
terms of order $1/\Lambda^4$ proportional to $L_{9L,9R}$, and terms
of order $1/\Lambda^4$ proportional to $L_{10}$ and $\beta_1$.

As mentioned before, we keep these terms together
to check our answer by taking the limit $L_{9L}=L_{9R}=0$.
This also allows us to cast our answer
in terms of $g_1^Z$, $\kappa_\gamma$ and $\kappa_Z$ which is convenient
for comparison with other papers in the literature.

It is amusing to note that the divergences generated by the operators
$L_{9L,9R}$ in the one-loop (order $1/\Lambda^4$) calculation of the
$Z \ra \overline{f} f$ widths can all be removed by the following
renormalization of the couplings in Eq.~\ref{oblique} (in the $M_t=0$ limit):
\beqn
\beta_1^r(\mu) &=& \beta_1 -
{\alpha \over \pi} {e^2\over 96 s_Z^4 c_Z^4}
{v^2 \over \Lambda^2}
\biggl[c_Z^2(1-20s_Z^2)L_{9L}+s_z^2(10-29c_Z^2)L_{9R}\biggr]
\biggl({1 \over \epsilon} + \log{\mu^2 \over M_Z^2}\biggr)
\nonumber \\
L^r_{10}(\mu)&=& L_{10} -
{\alpha \over \pi}{1\over 96 s_Z^2 c_Z^2}
\biggl[(1-24c_Z^2)L_{9L}+(32c_Z^2-1)L_{9R}\biggr]
\biggl({1 \over \epsilon} + \log{\mu^2 \over M_Z^2}\biggr)
\label{amusing}
\eeqn
This proves our assertion that our calculation to order ${\cal O}
(1/\Lambda^4)$ can be made finite by suitable renormalizations
of the parameters in Eq.~\ref{oblique}. However, we do not expect
this result to be true in general. That is, we expect that a
calculation of the one-loop contributions of the operators in
Eq.~\ref{lfour} to other observables will require counterterms
of order $1/\Lambda^4$. Thus, Eq.~\ref{amusing} does {\it not}
mean that we can place bounds on $L_{9L,9R}$ from global fits
to the parameters $S$ and $T$. Without performing a complete
analysis of the effective Lagrangian at order $1/\Lambda^4$
it is not possible to identify the renormalized parameters of
Eq.~\ref{amusing} with the ones corresponding to $S$ and $T$
that are used for global fits.

Combining all the results of Appendix~B into Eq.~\ref{full},
and keeping only terms
linear in $L_{9L,9R}$, we find after using Eq.~\ref{renorm}, and
our prescription Eq.~\ref{prescrip}:
\beqn
{\delta\Gamma^{L_9} \over \Gamma_0}&=&{\alpha^2 \over 24}
{1\over c_Z^4 s_Z^4}{v^2 \over \Lambda^2}
\log\biggl({\mu^2 \over M_Z^2}\biggr) \nonumber \\
&&\biggl\{ \biggl[L_{9L}(1-24c_Z^2)+L_{9R}(-1+32c_Z^2)\biggr]
{2r_f(l_f+r_f)\over l_f^2 +r_f^2}{c_Z^2 \over s_Z^2-c_Z^2}
\nonumber \\
&&+2\biggl[L_{9L}c_Z^2(1-20s_Z^2)+L_{9R}s_Z^2(10-29c_Z^2)\biggr]
\biggl(1+{2r_f(l_f+r_f)\over l_f^2 +r_f^2}
{c_Z^2 \over s_Z^2-c_Z^2}\biggr)\biggr\}
\nonumber \\
&&+{\alpha^2 \over 12}
\log\biggl({\mu^2 \over M_Z^2}\biggr)
{1+2c_Z^2 \over c_Z^4 s_Z^4 (l_b^2 +r_b^2)}
{v^2 \over \Lambda^2}
{M_t^2 \over M_Z^2}\biggl[L_{9R}s_Z^2-7L_{9L}c_Z^2\biggr]\delta_{fb}
\label{finalres}
\eeqn
The last term in Eq.~\ref{finalres} corresponds to the non-universal
corrections proportional to $M_t^2$ that are relevant only
for the decay $Z \ra \overline{b} b$.

Using, as before,
$\Lambda=2$~TeV, $\mu=1$~TeV we find $90\%$ confidence level bounds:
\beqn
\Gamma_e &\ra & -92 \leq L_{9L}+0.22L_{9R} \leq 47 \nonumber \\
\Gamma_\nu &\ra & -79 \leq L_{9L}+1.02L_{9R} \leq 170 \nonumber \\
R_h &\ra & -22 \leq L_{9L}-0.17L_{9R} \leq 16 \nonumber \\
\Gamma_Z &\ra & -22 \leq L_{9L}-0.04L_{9R} \leq 17
\label{rhonum}
\eeqn
We show these inequalities in Figure~3.
\begin{figure}[htb]
\centerline{\hfil\epsffile{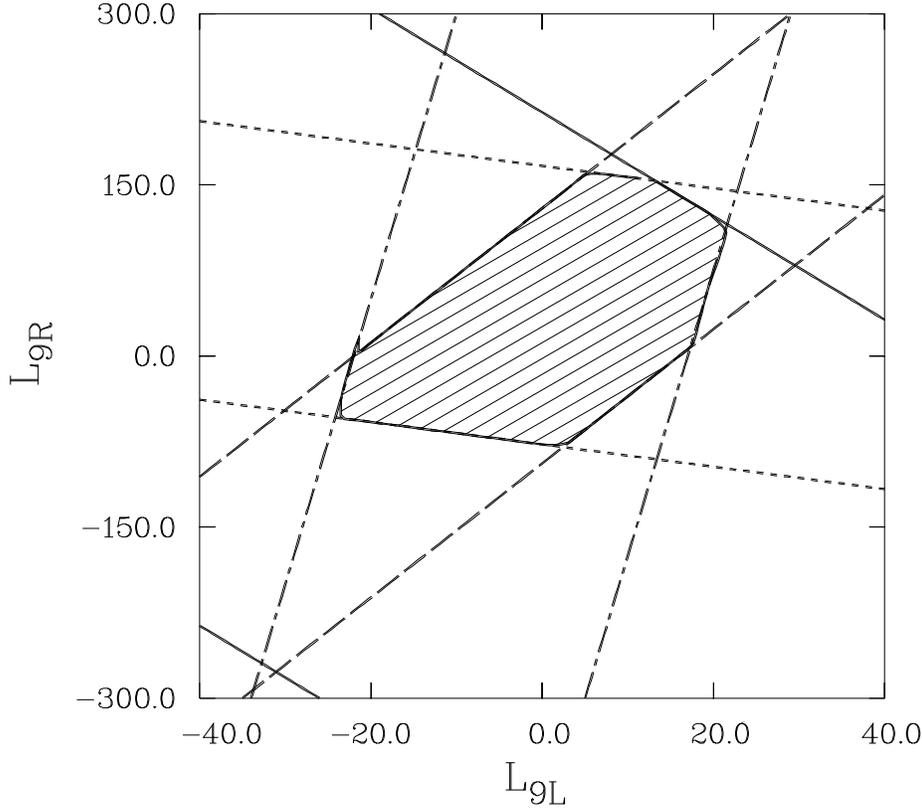}\hfil}
\caption[]{$90\%$ confidence level bounds on $L_{9L,9R}$ from the
$Z\rightarrow f {\overline f}$ partial widths, (Eq. 27).  The allowed
region is shaded.  The solid, dotted, dashed, and dot-dashed lines
are the bounds from $\Gamma_e$, $\Gamma_\nu$, $R_h$, and $\Gamma_Z$,
respectively.}
\end{figure}
If we bound one coupling at a time
we can read from Figure~3 that:
\beqn
-22 \leq &L_{9L}& \leq 16 \nonumber \\
-77 \leq &L_{9R}& \leq 94
\label{oneattime}
\eeqn
In a vector like model with $L_{9L}=L_{9R}$ we have
the $90\%$ confidence level bound:
\beq
-22<L_{9L}=L_{9R}< 18  .
\label{vectorbound}
\eeq

We can relate our couplings of Eq.~\ref{lfour} to the conventional
$g_1^Z$, $\kappa_\gamma$ and $\kappa_Z$ by identifying our unitary gauge
three gauge boson couplings with the conventional parameterization of
Ref.~\cite{hagi} as we do in Appendix~A. However, we must emphasize that
there is no unique correspondence between the two. Our framework assumes,
for example, $SU(2)_L\times U(1)_Y$ gauge invariance and this results in
specific relations between the three and four gauge boson couplings that are
different from those of Ref.~\cite{burgess} which assumes only electromagnetic
gauge invariance. Furthermore,
if one starts with the conventional parameterization of the three-gauge-boson
coupling and imposes $SU(2)_L\times U(1)_Y$ gauge invariance one does not
generate any additional two-gauge-boson couplings.
It is interesting to point out that within our formalism there are only
two independent couplings that contribute to three-gauge-boson couplings
($L_{9L,9R}$) but not to two-gauge-boson couplings (as $L_{10}$ does).
{}From this it follows that the equations for $g_1^Z$, $\kappa_\gamma$
and $\kappa_Z$ in terms of $L_{9L.9R,10}$ are not independent. In fact,
within our framework we have:
\beq
\kappa_Z = g_1^Z + {s_Z^2 \over c_Z^2}\biggl( 1 -\kappa_\gamma \biggr).
\label{depen}
\eeq
The same result holds in the formalism of Ref.~\cite{zeppenfeld}.

For the sake of comparison with the literature we translate the bounds
on $L_{9L}$ and $L_{9R}$ into bounds on $\Delta g_1^Z$, $\Delta \kappa_Z$
and $\Delta \kappa_\gamma$. For this exercise we set $L_{10}=0$.
We use $L_{9R}=0$ to obtain the bound
on $\Delta g_1^Z$. We then solve for $L_{9L}$ and $L_{9R}$  in terms of
$\Delta \kappa_Z$ and $\Delta \kappa_\gamma$, and bound each one of these
assuming the other one is zero. We obtain the $90\%$ confidence level
intervals:
\beqn
-0.08 < & \Delta g_1^Z& < 0.1\nonumber \\
-0.3 < & \Delta \kappa_Z & < 0.3\nonumber \\
-0.3 < & \Delta \kappa_\gamma & < 0.4  .
\label{ourdg}
\eeqn
Similarly, if there is a non-zero $L_{10}$, these couplings receive
contributions from it. Setting $L_{9L,9R}=0$, we find from Eq.~\ref{tennum}
the bounds: $-0.004 <  \Delta g_1^Z < 0.005$,
$-0.003 <  \Delta \kappa_Z  < 0.004$, and
$-0.009 <  \Delta \kappa_\gamma  < 0.007$.
These bounds are stronger by a factor of about 20, just as the bounds on
$L_{10}$, Eq.~\ref{tennum} are stronger by about a factor of 20 than the
bounds on $L_{9L,9R}$, Eq.~\ref{oneattime}. However, these really are bounds
on the oblique corrections introduced by $L_{10}$ (which also contributes
to three gauge boson couplings due to $SU(2)_L\times U(1)_Y$ gauge invariance).
It is perhaps more relevant to consider the couplings of operators without
tree-level self-energy corrections. This results in Eq.~\ref{ourdg}.

\subsection{Effects proportional to $M_t^2$}

As can be seen from Eq.~\ref{finalres}, the $Z\ra \overline{b} b$
partial width receives non-universal contributions proportional to
$M_t^2$. Within our renormalization scheme, the effects that correspond to
the minimal standard model do not occur. Our result corresponds entirely to
a new physics contribution of order $1/\Lambda^4$ proportional to
$L_{9L,9R}$. These effects have already been included to some extent in
the previous section when we compared the hadronic and total widths of the
$Z$ boson with their experimental values. In this section we isolate
the effect of the $M_t^2$ terms and concentrate on the $Z\ra \overline{b} b$
width. Keeping the leading non-analytic contribution, as usual, we find:
\beq
{\Gamma(Z\ra \overline{b} b) \over \Gamma_0}-1 =
{\alpha^2 \over 12}{1+2c_Z^2 \over c_Z^4 s_Z^4 (l_b^2 +r_b^2)}
{v^2 \over \Lambda^2}
{M_t^2 \over M_Z^2}\biggl[L_{9R}s_Z^2-7L_{9L}c_Z^2\biggr]
\log\biggl({\mu^2 \over M_Z^2}\biggr)
\label{nonuni}
\eeq
We use as before $\mu=1$~TeV, and we neglect the contributions
to the $Z \ra \overline{b} b$ width that are not proportional to
$M_t^2$. We can then place bounds on the anomalous couplings by
comparing with Langacker's result $\delta_{bb}^{new}=0.022 \pm
0.011$ for $M_t = 175 \pm 16$~GeV \cite{langap}. Bounding the
couplings one at a time we find the
$90\%$ confidence level intervals:
\beqn
-50 \leq &L_{9L}& \leq -4  \nonumber \\
 90 \leq & L_{9R}& \leq 1200 .
\label{dbblim}
\eeqn
Once again we find that there is much more sensitivity to
$L_{9L}$ than to $L_{9R}$. The fact that the $Z \ra \overline{b}b$
vertex places asymmetric bounds on the couplings is, of course,
due to the present inconsistency between the measured value and
the minimal standard model result. Clearly, the implication that
the couplings $L_{9L,9R}$ have a definite sign cannot be taken
seriously. A better way to read Eq.~\ref{dbblim} is thus:
$|L_{9L}|\leq 50$ and $|L_{9R}|\leq 1200$.

\section{Discussion}

Several studies that bound these ``anomalous couplings'' using the
LEP observables can be found in the literature. Our present study
differs from those in two ways: we have included bounds on some
couplings that have not been previously considered, $L_{1,2}$ and
we discuss the other couplings, $L_{9L,9R}$ within an $SU(2)_C
\times U(1)_Y$ gauge invariant formalism. We now
discuss specific differences with some of the papers found in the
literature.

The authors of Ref.~\cite{hervegas} obtain their bounds by regularizing the
one-loop integrals in $n$ dimensions, isolating the poles in $n-2$ and
identifying these with quadratic divergences. This differs from our
approach where we keep only the (finite) terms proportional to the
logarithm of the renormalization scale $\log\mu$. To find bounds, the
authors of Ref.~\cite{hervegas} replace the poles in $n-2$ with
factors of $\Lambda^2/M_W^2$. We believe that this leads to the
artificially tight constraints \cite{quaddiv}
on the anomalous couplings quoted
in Ref.~\cite{hervegas} ($2\sigma$ limits):
$-9.4\times 10^{-3} \leq \beta_2 \leq 2.2\times 10^{-2}$ and
$-1.5\times 10^{-2} \leq \beta_3 \leq 3.9\times 10^{-2}$. We translate
these into $90\%$ confidence level intervals:
\beqn
-1.0 \leq &L_{9R} & \leq 2.4 \nonumber \\
-1.6 \leq &L_{9L} & \leq 4.2
\label{vegas}
\eeqn
which are an order of magnitude tighter than our bounds. Conceptually,
we see the divergences as being absorbed by renormalization of other
anomalous couplings. As shown in this paper, the calculation of the
$Z\ra \overline{f}f$ can be rendered finite at order ${\cal O}(1/
\Lambda^4)$ by renormalization of $\beta_1$ and $L_{10}$. Thus, the
bounds obtained by Ref.~\cite{hervegas}, Eq.~\ref{vegas}, are really bounds on
$\beta_1$ and $L_{10}$. They embody the naturalness assumption
that all the coefficients that appear in the effective Lagrangian at
a given order are of the same size. Our formalism effectively
allows $L_{9L,9R}$ to be different from $L_{10}$.

The authors of Ref.~\cite{burgess} do not require that their effective
Lagrangian be $SU(2)_L\times U(1)_Y$ gauge invariant, and instead they are
satisfied with electromagnetic gauge invariance. At the technical level
this means that we differ in the four gauge boson vertices associated with the
anomalous couplings we study. It also means that we consider different
operators. In terms of the conventional anomalous three gauge boson couplings,
these authors quote $1\sigma$ results
$\Delta g_1^Z = -0.040 \pm 0.046$,
$\Delta\kappa_\gamma = 0.056 \pm 0.056$, and
$\Delta\kappa_Z = 0.004 \pm 0.042$.
These constraints are tighter than what we obtain from the contribution
of $L_{9L,9R}$ to $\Delta g_1^Z$, $\Delta\kappa_\gamma$ and
$\Delta\kappa_Z$, Eq.~\ref{ourdg}; they are weaker than what we obtain from the
contribution of $L_{10}$.

The authors of Ref.~\cite{zeppenfeld} require their effective Lagrangian to
be $SU(2)_L \times U(1)_Y$ gauge invariant, but they implement the symmetry
breaking linearly, with a Higgs boson field. The resulting power counting
is thus different from ours, as are the anomalous coupling constants. Their
study would be appropriate for a scenario in which the symmetry breaking
sector contains a relatively light Higgs boson. Their anomalous couplings
would parameterize the effects of the new physics not directly attributable
to the Higgs particle. Nevertheless, we can roughly compare our results to
theirs by using their bounds for the heavy Higgs case (case (d) in Figure~3
of Ref.~\cite{zeppenfeld}). To obtain their bounds they consider the case
where their couplings $f_B=f_W$ and $f_{WB}=0$ which corresponds to our
$L_{10}=0$, and $L_{9L}=L_{9R}$. For $M_t=170$~GeV they find the following
$90\%$ confidence level interval
$-0.05 \leq  \Delta\kappa_\gamma  \leq 0.12$, which we translate into:
\beq
-7.8 \leq  L_{9L}=L_{9R}  \leq 18.8
\label{zepp}
\eeq
This compares well with our bound
\beq
-22 \leq L_{9L}=L_{9R} \leq 18
\label{uscomp}
\eeq

Finally, if we look {\it only} at those corrections that are proportional
to $M_t^2 /M_W^2$ and that would dominate in the $M_t \ra \infty$ limit, we
find that they only occur in the $Z \ra \overline{b} b$ vertex. This means
that they can be studied in terms of the parameter $\epsilon_b$ of
Ref.~\cite{alta} or $\delta_{bb}$ of Ref.~\cite{langanew}. Converting our
result of Eq.~\ref{dbblim} to the usual anomalous couplings and
recalling that only two of them are independent at this order, we
find, for example:
\beqn
-0.28 \leq &\Delta g_1^Z& \leq -0.03 \nonumber \\
0.6 \leq &\Delta \kappa_\gamma& \leq 5.2
\label{compdbb}
\eeqn
This result is very similar to that obtained in Ref.~\cite{eboli}.

We now compare our results\footnote{
Our normalization of the $L_i$ is different from that
of Ref. \cite{boud,bdv,fls}.  We have translated their results into
our notation.}
 with bounds that future colliders are
expected to place on the anomalous couplings.
\begin{figure}[htb]
\centerline{\hfil\epsffile{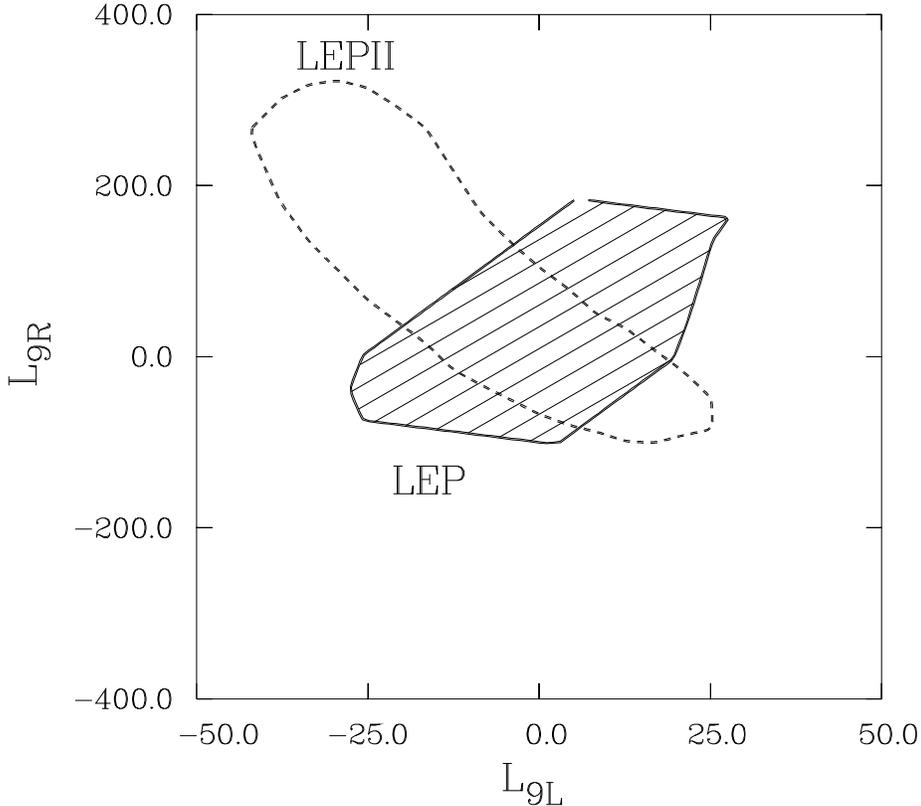}\hfil}
\caption[]{Comparison of the $95\%$ confidence level bounds from the
$Z$ partial widths (shaded region) with that obtainable at LEPII
with $\sqrt{s}=196~GeV$ and $\int {\cal L}=500~pb^{-1}$,
(dotted contour) \cite{boud}.}
\end{figure}
In Fig. 4, we compare
our $95\%$ confidence level bounds on $L_{9L}$ and $L_{9R}$ with those which
can be obtained at LEPII with $\sqrt{s}=196$~GeV and an integrated
luminosity of $500~pb^{-1}$ \cite{boud,bm2}.
  We find that LEP and LEPII are
sensitive to slightly different regions of the $L_{9L}$ and $L_{9R}$
parameter space, with the bounds from the two machines being of
the same order of magnitude.  The authors of Ref.~\cite{fls} find that
the LHC would place bounds of order $L_{9L} < {\cal O}(30)$ and a factor of
two or three worse for $L_{9R}$. We find, Eq.~\ref{oneattime},  that
precision LEP measurements
already provide constraints at that level.
We again emphasize our caveat that
the  bounds from LEP rely on naturalness arguments and are no substitute
for measurements in future colliders.

The limits presented here on the four point couplings
$L_{1}$ and $L_{2}$ are the first available for these couplings.
They will be measured directly at the LHC.  Assuming a coupling
is observable if it induces a $50\%$ change in the high momentum
integrated cross section, Ref.~\cite{bdv} estimated that the
LHC will be sensitive to $\mid L_{1}, L_{2}\mid \sim {\cal O}(1)$, which is
considerably stronger that the bound obtained from the $Z$
partial widths.

\section{Conclusions}

We have used an effective field theory formalism to place
bounds on some non-standard model gauge boson couplings.
We have assumed that the electroweak interactions are an
$SU(2)_L \times U(1)_Y$ gauge theory with an unknown,
but strongly interacting, scalar
sector responsible for spontaneous symmetry breaking.
Computing the leading contribution of each operator, and allowing
only one non-zero coefficient at a time, our $90~\%$ confidence level
bounds are:
\beqn
  -1.1 < &L_{10}^r(M_Z)_{new}& < 1.5 \nonumber \\
  -28 < &L_1& < 26  \nonumber \\
  -11 < &L_2& < 11  \nonumber \\
  -22 < &L_{9L}& < 16  \nonumber \\
  -77 < &L_{9R}& < 94.
\label{rescon}
\eeqn
Two parameter bounds on ($L_1,L_2$) and ($L_{9L},L_{9R}$) are
given in the text.  The bounds on $L_1,L_2$ are the first
experimental bounds on these couplings.  The bounds on $L_{9L}$
and $L_{9R}$ are of the same order of magnitude as those which
will be obtained at LEPII and the LHC.

\vspace {1.in}

\noindent{\bf Acknowledgements} The work of G. V. was supported in part
by a DOE OJI award. G.V. thanks the theory group at BNL for their
hospitality while part of this work was performed.
We are grateful to W. Bardeen, J.~F.~Donoghue, E. Laenen,
W. Marciano, A.~Sopczak, and A. Sirlin for useful discussions.
We thank P.~Langacker for providing us with his latest numbers.
We thank F.~Boudjema for providing us with the data file for
the LEPII bounds in Figure~4.

\appendix

\section{Three and Four Gauge Boson Couplings in Unitary Gauge}

It has become conventional in the literature to parameterize the
three gauge boson vertex $VW^+W^-$ (where $V=Z,\gamma$)
in the following way \cite{hagi}:
\beqn
{\cal L}_{WWV}&= &
-ie {c_Z\over s_Z} g_1^Z \biggl(
W_{\mu\nu}^{\dagger} W^{\mu}-W_{\mu\nu} W^{\mu~\dagger}\biggr) Z^\nu
-ie g_1^\gamma\biggl(
W_{\mu\nu}^{\dagger} W^{\mu}-W_{\mu\nu} W^{\mu~\dagger}\biggr) A^\nu
\nonumber \\ &&
-ie {c_Z\over s_Z} \kappa_Z
W_{\mu}^{\dagger} W_{\nu}Z^{\mu\nu}
-ie \kappa_\gamma W_{\mu}^{\dagger} W_{\nu}A^{\mu\nu}
\nonumber \\ & &
-e {c_Z \over s_Z} g_5^Z
\epsilon^{\alpha\beta\mu\nu}\biggl(
W_\nu^-\partial_\alpha W_\beta^+-W_\beta^+\partial_\alpha
W_\nu^-\biggr)Z_\mu \quad .
\label{conv}
\eeqn

Terms of the form $(\lambda_V/M_V^2)
W_{\rho\mu}^{\dag} W^\mu_\sigma V^{\sigma\rho}$
which are often included in the parameterization of the
three gauge boson vertex do not appear in our formalism to the
order we work.

For calculations to order $1/\Lambda^2$, it is most convenient to
diagonalize the gauge-boson self-energies as done in Ref.~\cite{holdom}.
This results in expressions for $g_1^Z$, $\kappa_Z$ and $\kappa_\gamma$
in terms of $L_{9L,9R,10}$ that we presented in Ref.~\cite{cdhv}.
For the present study, we do not keep the $L_{10}$ or $\beta_1$
terms as explained in the text. We thus use:
\beqn
g_1^Z&=&1+{e^2\over 2 c_Z^2 s_Z^2} L_{9L}{v^2\over\Lambda^2}\nonumber \\
g_1^\gamma&=& 1 \nonumber \\
\kappa_Z&=&1+ {e^2\over 2 s_Z^2c_Z^2} \biggl(L_{9L}c_Z^2
-L_{9R}s_Z^2\biggr){v^2\over \Lambda^2}\nonumber \\
\kappa_\gamma&=&1+{e^2 \over 2s_Z^2}
\biggl(L_{9L}+L_{9R}\biggr){v^2\over \Lambda^2}.
\label{unot}
\eeqn

The four gauge boson interactions derived from
Eqs.~\ref{lagt} and~\ref{lfour} after diagonalization of
the gauge boson self-energies can be written as:
\beqn
{\cal L}_{WWV_i V_j}&=&
C_{ij}
\biggl( 2 W^+\cdot W^- V_i\cdot V_j
-V_i \cdot W^+  V_j\cdot W^-
-V_j\cdot W^+ V_i\cdot W^-\biggr)\nonumber \\
&& +{e^4 \over s^4_Z} {v^2\over \Lambda^2}\biggl[ {1\over c_Z^2}
\biggl( L_1 W^+\cdot W^- Z \cdot Z +L_2 W^+\cdot Z W^-\cdot Z\biggr)
\nonumber \\
&&+(L_1+{L_2\over 2})(W^+\cdot W^-)^2 +{L_2\over 2}W^+\cdot W^+
W^-\cdot W^- \nonumber \\
&&+{1\over 4 c_Z^4}(L_1+L_2)(Z\cdot Z)^2\biggr]
\label{afbg}
\eeqn
where $V_i=\gamma, Z$ or $W^{\pm}$ and,
\beqn
C_{\gamma \gamma}&=& -e^2  \nonumber \\
C_{ZZ} &=& -e^2 {c_Z^2\over s_Z^2}(g_1^Z)^2\nonumber \\
C_{\gamma Z}&=&-e^2{c_Z\over s_Z}g_1^Z\nonumber \\
C_{WW}&=&-{e^2\over s_Z^2}\biggl(1+2 c_Z^2(g_1^Z-1)\biggr).
\label{fgbco}
\eeqn

\section{One-loop Results}

As explained in the text, we will only consider the tree-level effects
of $L_{10}$. This means that for the one-loop calculation to order
$1/\Lambda^4$ only $L_{9L,9R}$ appear in Eq.~\ref{unot}.
For the calculation to order
$1/\Lambda^2$ presented in this paper, we do not use the diagonal
basis, but rather obtain our results from the explicit factors of
$L_{10}$ and $\beta_1$ that appear in the following expressions.

The vector boson self energies can be written in the form:
\beq
-i \Pi_{VV}^{\mu\nu}(p^2)=
A_{VV}(p^2)g^{\mu\nu}+ B(p^2)p^{\mu}p^{\nu}.
\label{defse}
\eeq
We regularize in $n$ dimensions and keep only the poles of
ultraviolet origin. For the case of fermion loops we treat all
fermions as massless except the top-quark. We find:
\beqn
{A_{\gamma\gamma}(p^2)\over p^2}&=&-{\alpha\over 4 \pi\epsilon}
 \biggl\{
{p^2\over 3 M_W^2}-1+{\kappa_\gamma^2\over 12}{p^2\over M_W^4}(p^2-2M_W^2)
+{\kappa_\gamma\over M_W^2}(p^2-6M_W^2)\biggr\}
\nonumber \\
&&-{\alpha \over 12\pi s_Z^4\epsilon}\sum_f N_{Cf}r_f^2
+ 8 \pi \alpha {v^2 \over \Lambda^2} L_{10}
\nonumber \\
{A_{\gamma Z}(p^2)\over p^2}&=&{\alpha\over 4 \pi\epsilon}
\biggl({c_Z\over s_Z}\biggr)
\biggl\{g_1^Z(1-{p^2\over 3 M_W^2})-{\kappa_Z\kappa_\gamma p^2
\over 12 M_W^4}(p^2-2M_W^2)\nonumber \\
&-&{(g_1^Z\kappa_\gamma+\kappa_Z)\over 2 M_W^2}
(p^2-6 M_W^2)\biggr\}
\nonumber \\
&&+{\alpha \over 24 \pi s_Z^3 c_Z\epsilon}
 \sum_f N_{Cf} r_f \biggl(r_f+l_f\biggr)
+4\pi\alpha {c_Z^2-s_Z^2 \over s_Z c_Z}{v^2\over \Lambda^2}L_{10}
\nonumber \\
{A_{Z Z}(p^2)\over p^2}&=&{\alpha\over 4 \pi\epsilon}
\biggl({c_Z\over s_Z}\biggr)^2
\biggl\{g_1^{Z~2}(1-{p^2\over 3 M_W^2})-{\kappa_Z^2 p^2
\over 12 M_W^4}(p^2-2M_W^2)-{g_1^Z\kappa_Z\over  M_W^2}
(p^2-6 M_W^2)\biggr\}
\nonumber \\
&&-{\alpha \over 8 \pi s_Z^2 c_Z^2\epsilon}
\biggl[- {3 M_t^2\over p^2} +\sum_f {N_{Cf}\over 3}\biggl(
r_f^2+l_f^2\biggr)\biggr]
-2{M_Z^2 \over p^2}\beta_1-8\pi\alpha{v^2\over\Lambda^2}L_{10}
\label{avv}
\eeqn
These results can be compared with the unitary gauge results of
Degrassi and Sirlin in the standard model limit ($g_1^Z=\kappa_Z=
\kappa_\gamma=1$).  When the contribution of the standard model
Higgs boson is included, Eq.~\ref{avv} agrees with Ref.~\cite{ds}

For the renormalization of $G_F$, we need $A_{WW}(q^2=0)$,
the $W e \nu$ vertex evaluated at $q^2=0$, $\Gamma_{We\nu}$,
the box contribution
to $\mu \rightarrow e \nu {\overline \nu}$, $B_{\rm box}$,
and the charged lepton wavefunction renormalization, $Z_f$:
\beqn
A_{WW}(0)&=& {3 \alpha\over 16 \pi\epsilon} M_W^2
\biggl\{
\kappa_\gamma^2 + 2 + 2 \kappa_\gamma -{3\over s_Z^2}
(1-2 c_Z^2)\nonumber \\
& &+\biggl({1\over s_Z^2}\biggr)
\biggl[ \kappa_Z^2 \biggl( c_Z^2 + 1 - {1\over c_Z^2}\biggr)
+g_1^{Z~2}\biggl( c_Z^2 - 2 - c_Z^4\biggr)
\nonumber \\
&&
+2 \kappa_Z g_1^Z \biggl( c_Z^2 +1 \biggr)
- 6 c_Z^2 g_1^Z\biggr]\biggr\}
+{3\alpha\over 8 \pi s_Z^2 \epsilon}M_t^2
\nonumber \\
\Gamma_{W e \nu}(0)& =&
 {\cal A}_0 {3 \alpha\over 8 \pi \epsilon }
\biggl\{ \biggl(1+{1\over 2} \kappa_\gamma\biggr)
+{c_Z^2\over 2s_Z^2}
\biggl[g_1^Z\biggl(1-c_Z^2\biggr)
+\kappa_Z\biggl(1 - {1 \over c_Z^2}\biggr)\biggr]\biggr\}
\nonumber \\
{\cal A}_0&=& -{g\over 2 \sqrt{2}}{\overline e}\gamma^\mu
(1-\gamma_5) \nu \epsilon^W_\mu
\nonumber \\
B_{\rm box}&=&{3 \alpha\over 16 \pi \epsilon}\biggl({c_Z^2\over s_Z^2
}\biggr)(1+c_Z^2)+{\alpha\over 4 \pi\epsilon}
\nonumber \\
Z_f-1& =& -{\alpha r_f^2\over 16 s_Z^4 \pi \epsilon}
\label{intermediate}
\eeqn
For massless fermions in dimensional regularization there
is a cancellation between the ultraviolet and infrared divergent
contributions, responsible for the familiar result that their
wavefunction renormalization vanishes. We are isolating the
ultraviolet divergences only, so we obtain a contribution
to the fermion wavefunction renormalization.

\begin{figure}[htb]
\centerline{\hfil\epsffile{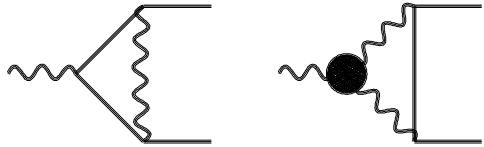}\hfil}
\caption[]{Diagrams contributing to $\delta l_f$.}
\end{figure}
The corrections to the $Z(p^2)  {\overline f} f$ vertex from the diagrams
shown in Fig. 5 (including in this term the wave function renormalization
for the external fermion) are:
\beqn
\delta l_f &=& (l_f-r_f){\alpha\over 4 \pi\epsilon}
\biggl({c_Z\over s_Z}\biggr)^2
 \biggl\{\kappa_Z{p^2\over M_W^2} \biggl( {1\over 2}+
{p^2\over 12 M_W^2}\biggr)
+ g_1^Z \biggl({5p^2\over 6 M_W^2}\biggr)\biggr\}
\nonumber \\
&&+{\alpha\over 16 \pi s_Z^2 \epsilon}
{M_t^2\over M_W^2}\biggl[r_t-4l_t+(\kappa_Z {p^2 \over M_Z^2}
+6  c_Z^2g_1^Z)+3L_b\biggr]\delta_{fb}
\label{moreinter}
\eeqn
When the wavefunction renormalization is included in the definition
of $\delta r_f$, we have $\delta r_f=0$ from the diagrams of Fig. 5.

The $Z$ wavefunction renormalization is given by:
\beqn
Z_Z-1&=&-{\alpha\over 8 \pi s_Z^2 c_Z^2\epsilon}\sum_f
{N_{Cf} \over 3}\biggl( r_f ^2+
l_f^2\biggr) -8\pi \alpha{v^2 \over \Lambda^2}L_{10}
\nonumber \\
&&+{\alpha\over 4 \pi\epsilon}
\biggl({c_Z\over s_Z}\biggr)^2
\biggl\{g_1^{Z~2}(1-{2\over 3 c_Z^2})-{\kappa_Z^2
\over 12 }({3\over c_Z^4}-{4\over c_Z^2})
-g_1^Z\kappa_Z
({2\over c_Z^2}-6 )\biggr\}
\label{wavefunct}
\eeqn


\begin{thebibliography}{999}

\bibitem{altanew}{G.~Altarelli, CERN-TH-7319/94.}

\bibitem{langanew}{P.~Langacker, UPR-0624T.}

\bibitem{pestak}{D.~C.~Kennedy and B.~W.~Lynn, \np{322}{1}{89};
M.~Peskin and T.~Takeuchi, \prl{65}{964}{90}.}

\bibitem{alta}{G.~Altarelli, and R.~ Barbieri, and F.~ Caravaglios,
\np{405}{3}{93}.}

\bibitem{longo}{T.~Appelquist and C.~Bernard, \prd{22}{200}{80};
A.~Longhitano, \np{188}{118}{81}.}

\bibitem{appel}{T.~Appelquist and G.-H.~Wu, \prd{48}{3235}{93}.}

\bibitem{siki}{P.~Sikivie, {\it et. al.}, \np{173}{189}{80}.}

\bibitem{boud} {F. Boudjema, {\it Proceedings of
Physics and Experiments with Linear
$e^+e^-$  Colliders}, ed. by F.~A.~Harris {\it et al.}, (1993),
p. 713, and references therein.}

\bibitem{georgi}{A.~Manohar and H.~Georgi, \np{234}{189}{84}.}

\bibitem{glas}{D.~Schaile, Plenary talk presented at the ICHEP-94 meeting,
Glasgow (1994).}

\bibitem{langap}{P.~Langacker, private communication.}

\bibitem{bernabeu}{The central value is from
J.~Bernabeu, A.~Pich and A.~Santamaria, \np{363}{326}{91}
for $M_t=170$~GeV. The errors are those given in Ref.~\cite{langa}.}

\bibitem{langa}{P.~Langacker, Precision Tests of the Standard Model, Lectures
given at TASI-92, Boulder Co. (1992).}

\bibitem{dvplep}{S.~Dawson and G.~Valencia, \prd{49} {2188}{94};
\pl{333}{207}{94} and erratum to appear.}

\bibitem{holdom}{B.~Holdom, \pl{258}{156}{91}.}

\bibitem{cdhv}{K.~Cheung, S.~Dawson. T.~Han and G.~Valencia,
UCD-94-6, NUHEP-TH-94-2, 1994.}

\bibitem{alls}{M.~Peskin and T.~Takeuchi, \prl, {65} {964} {90};
M.~Golden and L.~Randall, \np {361}  {3} {91};
W.~Marciano and J.~Rosner, \prl {65} {2963} {90};
D.~Kennedy and P.~Langacker, \prl {65} {2967} {90};
\prd {44} {1591} {91};
 G.~Altarelli and R.~Barbieri,
\pl {253} {161} {90}; B.~Holdom and J.~Terning, \pl
{247} {88} {90};
G.~Altarelli, R.~Barbieri, and S. Jadach, \np {369}{3}{92}.}

\bibitem{herrero}{M.~Herrero and E.~Morales, \np{418}{431}{94}.}

\bibitem{bdv}{J.~Bagger, S.~Dawson and G.~Valencia, \np{399}{364}{93}.}

\bibitem{fls}{A.~Falk, M.~Luke and E.~Simmons, \np{365}{523}{91}.}

\bibitem{hagi}{K.~Hagiwara, {\it et. al.}, \np{282}{253}{87}.}

\bibitem{burgess}{C.~P.~Burgess, {\it et. al.}, McGill-93/14 (1993).}

\bibitem{zeppenfeld}{K.~Hagiwara, {\it et. al.}, \pl{283}{353}{92}.}

\bibitem{hervegas}{P.~Hernandez and F.~Vegas, \pl{307}{116}{93}.}

\bibitem{quaddiv}{M.~Einhorn and J.~Wudka, UM-TH-92,25;
C.~Burgess and D.~London, \prl{69}{3428}{92}; \prd{48}{4337}{93}.}

\bibitem{eboli} {O.~Eboli, {\it et. al.}, MAD/PH/836.}

\bibitem{bm2}{G.~Gounaris {\it et.al.} in {\it Proc. of the Workshop on $e^+
e^-$ Collisions at $500~GeV$: The Physics Potential}, DESY-92-123B.
p. 735, ed. P. Zerwas;
M.~Bilenky {\it et.al.} BI-TP 92/44 (1993).}

\bibitem{ds} {G.~ DeGrassi and A. ~Sirlin, \np{383} {73}{92}.}

\end{thebibliography}
\end{document}